\documentclass[aps,prc,showpacs,superscriptaddress,floatfix]{revtex4-1}    

\usepackage[dvips]{graphicx}
\usepackage{color}
\usepackage{amssymb}
\usepackage{amsmath}
\usepackage{braket}
\usepackage{hyperref}
\usepackage{bm}

\newcommand{\nc}{\newcommand}           
\nc{\vc}[1]     {\mbox{\boldmath $#1$}} 
\nc{\mapleft}[1]{                       
 \smash{\mathop{                        %
  \hbox to 0.90cm{\rightarrowfill} }\limits_{#1}}}
\nc{\figwidth}{0.45}                    

\nc{\mydraft}	{\setlength{\topmargin}{-1.5cm}}
\mydraft   

\begin{document}
\title{Contact representation of short range correlation in light nuclei studied 
by the High-Momentum Antisymmetrized Molecular Dynamics}

\author{Qing Zhao} \email[]{zhaoqing91@outlook.com}
\affiliation{School of Physics and Key Laboratory of Modern Acoustics, Institute
of Acoustics, Nanjing University, Nanjing 210093, China}

\author{Mengjiao Lyu} \email[]{mengjiao@rcnp.osaka-u.ac.jp}
\affiliation{Research Center for Nuclear Physics (RCNP), Osaka University, Osaka
567-0047, Japan}
    
\author{Zhongzhou Ren} \email[]{corresponding author: zren@tongji.edu.cn}
\affiliation{School of Physics Science and Engineering, Tongji University,
Shanghai 200092, China}  

\author{Takayuki Myo} \email[]{takayuki.myo@oit.ac.jp}
\affiliation {General Education, Faculty of Engineering, Osaka Institute of
Technology, Osaka, Osaka 535-8585, Japan} \affiliation {Research Center for
Nuclear Physics (RCNP), Osaka University, Osaka 567-0047, Japan}

\author{Hiroshi Toki} 
\affiliation {Research Center for Nuclear Physics (RCNP), Osaka University,
Osaka 567-0047, Japan}

\author{Kiyomi Ikeda} 
\affiliation{RIKEN Nishina Center, Wako, Saitama 351-0198, Japan}  

\author{Hisashi Horiuchi} 
\affiliation {Research Center for Nuclear Physics (RCNP), Osaka University,
Osaka 567-0047, Japan}

\author{Masahiro Isaka} 
\affiliation{Hosei University, 2-17-1 Fujimi, Chiyoda-ku, Tokyo 102-8160, Japan}
  
\author{Taiichi Yamada} 
\affiliation{Laboratory of Physics, Kanto Gakuin University, Yokohama 236-8501,
Japan}

\begin{abstract}
The high-momentum antisymmetrized molecular dynamics (HMAMD) is a new promising
framework with significant analytical simplicity and efficiency inherited from
its antisymmetrized molecular dynamics in describing the high momentum
correlations in various nuclear states. In the aim of further improving the
numerical efficiency for the description of nucleon-nucleon correlation, we
introduce a new formulation by including a new Gaussian weighted basis of high
momentum pairs in the HMAMD wave function, with which very rapid convergence is
obtained in numerical calculation. It is surprising that the very high-momentum
components in the new HMAMD basis are found to be almost equivalent to the
contact representation of the nucleon-nucleon pairs with very small nucleon-nucleon
distance. The explicit formulation for the contact term significantly
improves the numerical efficiency of the HMAMD method, which shows the
importance of the contact correlation in the formulation of light nuclei.
\end{abstract}

\maketitle

\section{Introduction}
In recent years, significant advances have been obtained in the {\it ab initio}
studies of many body physics in atoms, molecules, nuclei and hadrons. In nuclear
physics, the {\it ab initio} theories provide the correct descriptions of the
nucleon-nucleon ($NN$) correlations induced by the $NN$ interactions, which are
usually unfeasible in other nuclear models. In the bare $NN$ interactions, which
is determined phenomenologically by fitting $NN$ scattering observables, there
exist strong short-range repulsion and strong tensor interaction inducing $D$-wave
components in nuclei \cite{jastrow55, wiringa95, pieper01, feldmeier98}. To
describe the complex $NN$-correlations in nuclear many-body problems, Jastrow
constructs a trial function which is a product of correlation functions for each
$NN$ pair \cite{jastrow55}. This function is used in the Green's Function Monte
Carlo (GFMC) studies of finite nuclei as an input trial function
\cite{pieper01}. Another important approach is the Unitary Correlation Operator
Method (UCOM), where short-range correlations are described by unitary
transformations in exponential form \cite{feldmeier98,neff03,myo09}.

In recent works, a new variational approach is proposed for the {\it ab initio}
calculation of light nuclei, namely the tensor-optimized antisymmetrized
molecular dynamics (TOAMD) \cite{myo15,myo17a,myo17b,myo17c,myo17d}. In this
method, the correlation functions are introduced and multiplied to reference
wave function of Antisymmetrized Molecular Dynamics (AMD). Instead of the
product form in Jastrow approach, the power series expansion for the correlation
functions is used in TOAMD. It is found that the TOAMD wave function provides
better numerical accuracy comparing to the Variational Monte Carlo (VMC) method,
which uses Jastrow type correlation functions \cite{myo17d}. In {\it ab initio}
calculations using the AV8$^\prime$ bare interaction, the GFMC results of
$^{3}$H and $^{4}$He are correctly reproduced with the TOAMD method including
first two orders of power expansion \cite{pieper01}. Furthermore, similar
correlation functions have been formulated for Fermi sphere \cite{yamada18} and
successfully applied to the nuclear matter calculation \cite{yamada18b}. The
detailed formation of the TOAMD approach can be found in Refs.~\cite{myo15,
myo17c}.

On the basis of the same power expansion scheme for the correlation functions,
another description of $NN$ correlations has been invented by introducing
high-momentum (HM) nucleon-nucleon pairs into the AMD reference states, namely
the ``high-momentum antisymmetrized molecular dynamics (HMAMD)" \cite{myo17e,
myo18}. In calculations using the AV4$^\prime$ central interaction where the
strong short-range repulsion is included, the HMAMD is found to provide the same
numerical accuracy as the TOAMD and GFMC methods \cite{myo18}. The tensor
correlation is also nicely described in the HMAMD approach, in which two-particle two-hole excitations are completely described \cite{myo17e}. The
major advantage of the HMAMD approach is its analytical simplicity comparing to
other {\it ab initio} frameworks which makes it a very promising method to be
extended to general nuclear systems. However, the basis number of HMAMD method depends on the number of HM pairs and is
relatively large, as comparing to the TOAMD method \cite{lyu18b}. To reduce the
model space, in Refs.~\cite{lyu18,lyu18b} a hybridized approach
``tensor-optimized high-momentum antisymmetrized molecular dynamics (TO-HMAMD)"
is proposed to consider the balance between the analytical and numerical
efforts. On the other hand, the HMAMD method itself can become even more
successful if its efficiency in describing the $NN$ correlations can be further
improved.

In this work, we introduce a new formulation for the high-momentum part of the
short range correlation in the framework of HMAMD.  We calculate the $^{3}$H nucleus
with this new formation and show the importance of contact correlation of
$NN$ pairs, where two nucleons contact at almost the same spatial position. Using
an effective wave function, the nuclear contacts have been extracted from
experimental observables in Ref.~\cite{weiss18}. In {\it ab initio}
calculations, the contact correlation is mathematically included in the model
space, but it is generally treated as a component of short-range correlation. In
this work, we show that the contact correlation appears as a characteristic feature in the variational calculation of light nuclei for the treatment of the short-range correlation. Furthermore, the numerical efficiency of HMAMD can be significantly improved in this work because of the explicit formulation of the contact correlations.

This paper is organized as follows. In Sec.~\ref{sec:waveFunction}, we introduce
our new formulation of the HMAMD wave function. In Sec.~\ref{sec:results}, this new formulation is applied to the $^{3}$H nucleus using the AV4$^\prime$ interaction.  In Sec.~\ref{sec:zero}, we show the importance of the contact correlation using an analytical derivation.  The last Sec.~\ref{sec:conclusion} contains conclusions.

\section{Formulation}
\label{sec:waveFunction}
We start writing the AMD (Antisymmetrized Molecular Dynamics) basis to construct
the wave function for a nucleus of $A$ nucleons:
\begin{equation}
\ket{\Phi_{\text{AMD}}} = \ket{\mathcal{A}\{\phi_1(\mathbf{r}_1)...\phi_A(\mathbf{r}_A)\}}
~,
\end{equation}
where the single-nucleon wave functions $ \phi(\mathbf{r})$ are expressed in the
Gaussian wave packet with the range parameter $\nu$ and the centroid position
$\mathbf{Z}$ multiplied by the spin-isospin wave function $\chi_{\tau,\sigma}$.
\begin{equation}
\phi(\mathbf{r}) = (\frac{2\nu}{\pi})^{3/4}e^{-\nu(\mathbf{r}-\mathbf{Z})^2}\chi_{\tau,\sigma}~.
\end{equation}
For the s-shell nuclei $^3$H and $^4$He, the centroids are optimized to be
$\mathbf{Z} = \mathbf{0}$, and the corresponding single nucleon wave functions
reduce to the s-wave states. This is the result of the energy variation in TOAMD
\cite{myo17c, myo17d}.

Then we introduce the HMAMD bases as in Refs.~\cite{myo17e,myo18,lyu18,lyu18b}.
For the ground states of s-shell nuclei, the real parts of the centroid position
$\mathbf{Z}$, which represent the spatial positions of the Gaussian centroid,
are set zero. The imaginary part of $\mathbf{Z}$ represents the mean momentum of
Gaussian as
\begin{equation}
\langle\mathbf{p}\rangle=2\hbar\nu \textrm{Im} (\mathbf{Z}). 
\end{equation}
By assigning the imaginary components to the Gaussian centroids of two nucleons,
$\mathbf{Z}_1$ and $\mathbf{Z}_2$ with the same magnitude in opposite
directions, we introduce the high-momentum excitation of di-nucleon pairs
\cite{itagaki18} into AMD basis as
\begin{equation}
\mathbf{Z}_1 = i\mathbf{D}, \qquad  \mathbf{Z}_2 = -i\mathbf{D}.
\end{equation} 
which is the HMAMD basis $\ket{\Phi_{\text{HMAMD}}(\mathbf{D}_{\alpha})}$ with a
shift parameter $\mathbf{D}_{\alpha}$. With these imaginary shifts, the
correlation between a two-nucleon pair is introduced to the AMD basis in the 
momentum space, while the center-of-mass momentum remains to be zero. For the
$^3$H nucleus, we prepare three kinds of high-momentum pairs for different
spin-isospin channels as follows:
\begin{equation}\label{eq:pair}
\text{p}_\uparrow \, \text{and} \, \text{n}_\uparrow, \quad \text{p}_\uparrow \, \text{and} \, \text{n}_\downarrow, \quad \text{n}_\uparrow \, \text{and} \, \text{n}_\downarrow.
\end{equation}
When only a single high-momentum pair is included in each HMAMD basis, we name the
corresponding method ``single HMAMD'' \cite{myo17e}. When bases with double
high-momentum pairs are included in model space, the corresponding method is
named as ``double HMAMD'', which could reproduce nicely the results of GFMC in
the calculation of the $^{3}$H nucleus \cite{myo18}. In this study, we adopt the
single HMAMD approach to study the short-range correlations between nucleons.

In our previous studies \cite{myo17e,myo18}, the total wave function is
formulated as a linear combination of different HMAMD basis states in the
following form,
\begin{equation}\label{eq:dis}
\ket{\Psi_{\text{AMD+Dis}}} = C_0\ket{\Phi_{\text{AMD}}}
+\sum_{\alpha}C_{\alpha}\ket{\Psi_{\text{HMAMD}}(\mathbf{D}_{\alpha})}.
\end{equation}
The first term in Eq. (\ref{eq:dis}) is the AMD basis which describes the
(0s)$^3$ configuration. The second term is the superposition of the HMAMD bases with
sets of the discrete shift parameters $\{\mathbf{D}_{\alpha}\}$, which we name
as ``discrete HMAMD basis'' (AMD+Dis). In this term, the label $\alpha$ denotes the shift
parameter $\mathbf{D}_{\alpha}$ and spin-isospin components in Eq. (\ref{eq:pair}) to identify the HMAMD basis
states. The total energy and the coefficients $C_{\alpha}$ are obtained by
solving the eigenvalue problem with respect to the Hamiltonian. Theoretically,
superposing large number of discrete HMAMD basis is sufficient to obtain the
exact solution, as in Refs.~\cite{myo17e,myo18}. 

In the present study, we propose a new kind of wave function as a linear
combination of different HMAMD basis states, 
\begin{equation}\label{eq:total}
\ket{\Psi_{\text{AMD+Dis+Gau}}} = C_0\ket{\Phi_{\text{AMD}}}
+\sum_{\alpha}C_{\alpha}\ket{\Psi_{\text{HMAMD}}(\mathbf{D}_{\alpha})}
+C_\beta \int d\mathbf{D} e^{-\frac{\mathbf{D}^2}{\beta}}\ket{\Psi_{\text{HMAMD}}(\mathbf{D})}.
\end{equation}
In order to investigate analytically the short-range correlation between two
nucleons in the high-momentum pair, we introduce additionally the third term 
\begin{equation}\label{eq:int}
  \int d\mathbf{D} e^{-\frac{\mathbf{D}^2}{\beta}}\ket{\Psi_{\text{HMAMD}}(\mathbf{D})}
\end{equation}
of integration form in Eq. (\ref{eq:total}), which is named as ``Gaussian HMAMD
basis''. The Gaussian parameter $\beta$ is treated as the variational parameter.
This term has only two parameters, {\it i.e.}, the Gaussian parameter $\beta$
and the superposition coefficient $C_\beta$, which provide a crystal description
of the $NN$-correlations because of its simplicity in analytical derivations.

To show the effects of this Gaussian HMAMD bases, we also perform calculations
using the superposed ``AMD+Gau" wave function as the reduced form of Eq.
(\ref{eq:total}).
\begin{equation}
\ket{\Psi_{\text{AMD+Gau}}} = C_0\ket{\Phi_{\text{AMD}}}
+C_\beta\int d\mathbf{D} e^{-\frac{\mathbf{D}^2}{\beta}}\ket{\Psi_{\text{HMAMD}}(\mathbf{D})}.
\end{equation}

Finally, all the total wave functions are projected on to the eigenstates of the
angular momentum $J$ by using the angular momentum projection \cite{schuck80}:
\begin{equation}\label{eq:ap}
  \begin{split}
  \left| \Psi^{JM}\right\rangle
  =\frac{2J+1}{8\pi^{2}}\int d \Omega D^{J*}_{MK}(\Omega)\hat R (\Omega)
      \ket{\Psi}~.
  \end{split}
\end{equation}

The Hamiltonian adopted for the $^{3}$H nucleus is the Argonne V4$^\prime$ $NN$
central potential, which includes a strong short-range repulsion. The range
parameter $\nu$ is set to be 0.22 $\text{fm}^{-2}$ for the single-nucleon wave
functions as in Ref.~\cite{myo18}. To simplify the calculation, we consider
the imaginary shifts $\mathbf{D}$ only in the $z$ direction and take the magnitude of
$D_z$ in step of 1 fm for the discrete HMAMD bases. The numerical
integration in the Gaussian HMAMD basis in Eq. (\ref{eq:int}) is performed by
using the Monte Carlo technique. We also put the truncation at $D_z=18$ fm for this
integration, which is large enough for the numerical convergence.

\section{Results}
\label{sec:results}
We compare the numerical results of $^3$H, calculated by using the wave
functions of AMD, AMD+Gaussian, AMD+Discrete+Gaussian, and AMD+Discrete. We list
the corresponding total energy and the discrete HMAMD basis included in Table
\ref{table:result}. The Gaussian parameter $\beta$ is optimized for each wave
function, respectively. The optimized value is $2.27$ fm for the AMD+Gaussian wave
function and $2.22$ fm for the AMD+Discrete+Gaussian wave function. 

We show in Table \ref{table:result}  the energy of $^{3}$H corresponding to
the AMD+Discrete+Gaussian wave function $-7.38$ MeV in this work, which converges
with the result of $-7.40$ MeV of the AMD+Discrete wave function in
Ref.~\cite{myo18}. It means that the superposition of the Gaussian form basis with the
AMD basis and the discrete HMAMD basis is actually describing the same wave
function as the AMD+Discrete wave function used in our previous works.
Furthermore, in this new AMD+Discrete+Gaussian wave function, the discrete HMAMD
basis with the magnitude of $D_z$ from 1 to 6 fm are enough for the convergence
of calculation, which means that the discrete HMAMD basis with $D_z$ larger than 6
fm can be replaced by the single Gaussian HMAMD basis. Hence, in this new
AMD+Discrete+Gaussian wave function, the number of bases necessary to converge
to the exact solution is significantly reduced from 15 to 8, which could
simplify, in large scale, future calculations using the HMAMD framework.

\begin{table*}[htbp]
  \begin{center}
    \caption{Results of $^3$H corresponding to the wave functions of AMD,
    AMD+Gaussian, AMD+Discrete+Gaussian, and AMD+Discrete. The AV4$^\prime$
    potential is adopted for the $NN$ interaction. The second row includes the
    total energies of $^3$H calculated by using each wave function. The third
    row represents the shift parameters $D_{z}$ of the superposed discrete HMAMD
    basis. The forth row includes the number of bases in each wave
    function.\label{table:result}}
    \vspace{2mm}
 \begin{tabular*}{15cm}{ @{\extracolsep{\fill}} l c c c c c}
    \hline
    \hline
 &AMD     &AMD+Gau &AMD+Dis+Gau   &AMD+Dis \cite{myo18}\\
    \hline
Energy (MeV)           &$15.6$        &$-1.76$        &$-7.38$
&$-7.40$ \\
Discrete bases (fm) &$-$     &$-$            &$1,2,\dots , 6$   &$1,2,\dots ,
14$ \\
Total number of bases  &$1$             &$2$               &$8$           &$15$
\\
    \hline
    \hline
  \end{tabular*}
  \end{center}
\end{table*} 

\begin{figure}[htbp]
  \centering
  \includegraphics[width=\figwidth\textwidth]{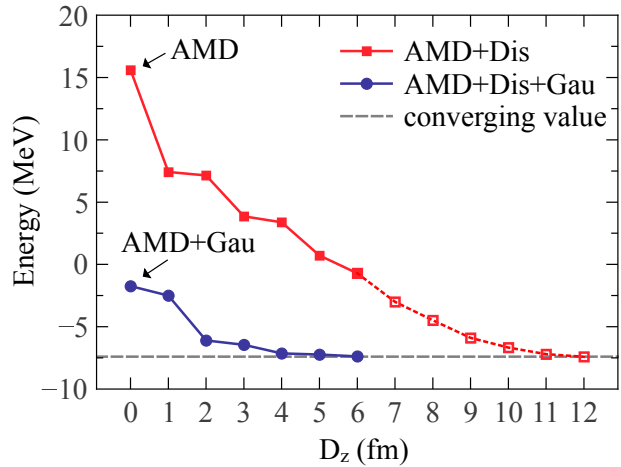}
  \caption{\label{fig:HMAMD_Gau}
  Energy curves for $^3$H of the AMD+Discrete (red line) and
  AMD+Discrete+Gaussian (blue line) calculations with respect to the successive
  addition of discrete HMAMD bases with the imaginary shift $D_z$. The dotted part
  of red line are the energies in AMD+Discrete with $D_z>6$, which are not
  necessary in the AMD+Discrete+Gaussian calculation. The first points of red and
  blue curves with $D_z=0$ correspond to the AMD and AMD+Gaussian results,
  respectively.}
\end{figure}

To show the effect of the Gaussian HMAMD basis explicitly, we present in Fig.
\ref{fig:HMAMD_Gau} the energy curve with respect to the successive addition of 
discrete HMAMD bases with the imaginary shift $D_z$ from both AMD and AMD+Gaussian
wave functions.  In this figure, the red curve shows the AMD+Discrete
calculation, where the discrete HMAMD bases are added successively increasing
$D_z$ to the AMD basis, and the energy convergence is obtained at around $D_z =
12$ fm. For the blue line, the first point corresponds to the energy by
superposing the AMD basis and the Gaussian HMAMD basis, and in the following
points additional discrete HMAMD bases are added successively. It is found that
the energy of $^3$H in each wave function converges to the same value of $-7.40$
MeV, which is obtained much faster with the Gaussian HMAMD basis with small number
of discrete HMAMD basis. This result indicates that very high-momentum
components with $D_z > 6$ fm in discrete HMAMD bases can be well described by
the simple Gaussian HMAMD basis.

\begin{figure}[htbp]
   \centering
   \includegraphics[width=\figwidth\textwidth]{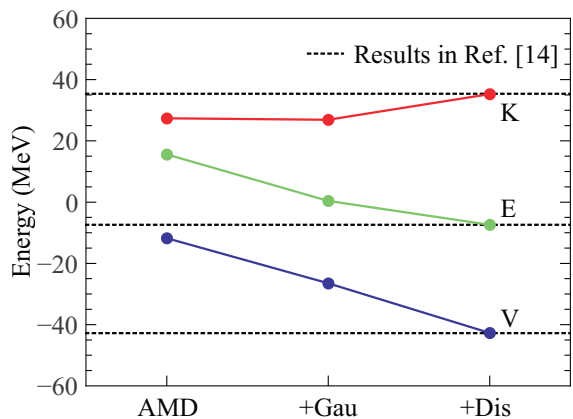}
   \caption{\label{fig:HM_KEV}
   The total energy $E$ and the Hamiltonian components of $^3$H by adding Gaussian
   form basis (+Gau) and the discrete HMAMD basis (+Dis) successively to the AMD
   basis. The symbols K and V indicate the kinetic and central interaction
   energies, respectively. Dashed lines represent the corresponding results of
   AMD+Discrete calculation in Ref.
   \cite{myo18}.}
 \end{figure}

In Fig. \ref{fig:HM_KEV}, we show the contribution to the total energy and
Hamiltonian components from the AMD basis and further successive addition of the
Gaussian HMAMD basis and discrete HMAMD bases. We obtain very nice agreement for
the total energy and each Hamiltonian component between the
AMD+Gaussian+Discrete results and the AMD+Discrete results from Ref.
\cite{myo18}. In the blue curve, a large contribution in the central interaction
energy is observed by the addition of the Gaussian HMAMD basis, which indicates the
importance of the Gaussian term in describing the short-range correlation. It is
also interesting that in the red curve the kinetic energy remains the same when
adding the Gaussian HMAMD basis, showing that the $NN$ correlation described by
Gaussian HMAMD basis, does not contribute to the kinetic energy. 
In the summary of these results, we clearly confirm the importance and
efficiency of the Gaussian HMAMD basis in short-range correlation.

\section{Contact representation of short range correlation}
\label{sec:zero}
In previous results, it is very surprising that the simple Gaussian HMAMD basis
in Eq. (\ref{eq:int}) actually provides an important contribution to the
short-range correlation. In order to understand the property of the nucleon-nucleon
correlation in this term, we investigate the analytical property of this
Gaussian HMAMD basis. We start with the spatial part of the single particle wave
function in HMAMD bases,
\begin{equation}
   \begin{split}
\phi(\mathbf{r}\pm i\mathbf{D}) = (\frac{2\nu}{\pi})^{3/4}e^{-\nu(\mathbf{r}\pm i\mathbf{D})^2},
   \end{split}
\end{equation}
where the real parts of centroid position $\mathbf{Z}$ are optimized to be zero
as in Eq. (2). Then, the high-momentum pair in the HMAMD basis can be
constructed as
\begin{equation}
   \begin{split}
\phi_1(\mathbf{r}_1-i\mathbf{D})\phi_2(\mathbf{r}_2+i\mathbf{D}),
   \end{split}
\end{equation}
where antisymmetrization is ignored for simplicity. In the Gaussian HMAMD basis,
we perform the numerical integration of Gaussian form for the shift parameter
$\mathbf{D}$:
\begin{equation}\label{eq:before-simplify}
\int d\mathbf{D} e^{-\mathbf{D}^2/\beta}\phi_1(\mathbf{r}_1-i\mathbf{D})\phi_2(\mathbf{r}_2+i\mathbf{D}) \propto e^{-\nu(\mathbf{r}_1^{2}+\mathbf{r}_2^2)} \int d\mathbf{D} e^{2\nu i \mathbf{D}(\mathbf{r}_1-\mathbf{r}_2)+(2\nu-\frac{1}{\beta})\mathbf{D}^2}.
\end{equation}
In the right hand side of Eq.~(\ref{eq:before-simplify}), the coefficient
$(2\nu-\frac{1}{\beta})$ for $\mathbf{D}^{2}$ in the exponent depends on the
parameter $\nu$, which is 0.22 fm$^{-2}$, and the variational optimized parameter $\beta$,
which controls the Gaussian distribution of $D$, namely the momentum. In the
results of the variational calculation, the optimized parameter $\beta$ is
obtained to be 2.22 $\text{fm}^2$, and hence we have approximately the following
numerical relation
\begin{equation}
  \nu\times\beta \approx 0.5,  
\end{equation}  
with which the integration in Eq.~(\ref{eq:before-simplify}) reduces to
\begin{equation}\label{eq:simplified}
   \begin{split}
\int d\mathbf{D} e^{-\frac{\mathbf{D}^2}{\beta}}\phi_1(\mathbf{r}_1-i\mathbf{D})\phi_2(\mathbf{r}_2+i\mathbf{D}) & \propto e^{-\nu(\mathbf{r}_1^{2}+\mathbf{r}_2^2)} \int d\mathbf{D} e^{2\nu i \mathbf{D}(\mathbf{r}_1-\mathbf{r}_2)} \\
&\propto \delta(\mathbf{r}_1-\mathbf{r}_2)\phi(\mathbf{r}_{1})\phi(\mathbf{r}_{2}).\\
   \end{split}
\end{equation}
This simple expression of the delta function shows a quite important result: the
Gaussian HMAMD basis describes the $NN$ correlation with the relative position
of two nucleons being zero, indicating spatial contact of two nucleons. This
result represents the contact correlation, as a very specific component of the
short-range correlation between nucleons. 

\begin{figure}[htbp]
  \centering
  \includegraphics[width=\figwidth\textwidth]{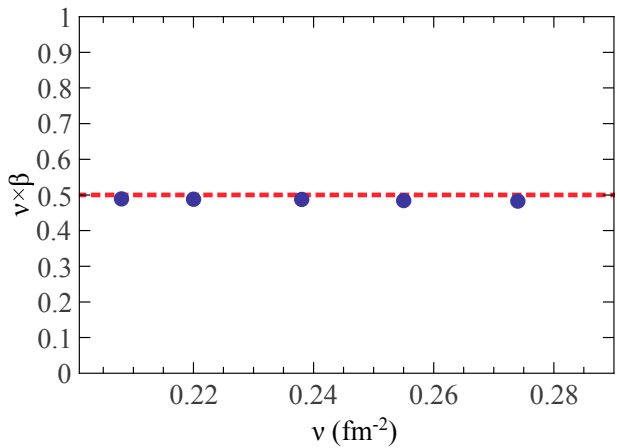}
  \caption{\label{fig:nuPlusBeta}
  The figure with different values of parameter $\nu$ and the values of
  $\nu\times\beta$. The corresponding values for parameter $\beta$ are
  variationally optimized and presented as the blue dots. The red dotted line
  denotes the $0.5$ value.}
\end{figure}
In order to further check the numerical relation ``$\nu\times\beta = 0.5$'', we
perform calculations of AMD+Discrete+Gaussian with various choices of parameter
$\nu$ and obtain the corresponding optimized $\beta$ parameter with variational
calculation, as shown in Fig. \ref{fig:nuPlusBeta}.  It is clearly observed that
for different choices of parameter $\nu$, the numerical relation
``$\nu\times\beta = 0.5$'' persists, which shows that this numerical relation is
not a parameter dependent coincidence. This coincidence strongly supports the
existence of the contact correlation in our Gaussian HMAMD basis. 

Meanwhile, we observe slight but finite difference between optimized values of
$\nu\times\beta$ and $0.5$ in Fig.~\ref{fig:nuPlusBeta}, which indicates that the
nucleon-nucleon distance in the contact correlated pair is not exact zero and
the $\delta$ function form in Eq.~(\ref{eq:simplified}) originates from the
approximation of $\nu\times\beta\approx 0.5$. Without this approximation, the
exact solution for the integration in Eq.~(\ref{eq:before-simplify}) can be
obtained, as introduced in Ref.~\cite{lyu18b},
\begin{equation}\label{eq:exact}
  \begin{split}
\int d\mathbf{D} e^{-\mathbf{D}^2/\beta}\phi_1(\mathbf{r}_1-i\mathbf{D})\phi_2(\mathbf{r}_2+i\mathbf{D}) & \propto e^{-\nu(\mathbf{r}_1^{2}+\mathbf{r}_2^2)} \int d\mathbf{D} e^{2\nu i \mathbf{D}(\mathbf{r}_1-\mathbf{r}_2)+(2\nu-\frac{1}{\beta})\mathbf{D}^2} \\
& \propto 
  e^{-\frac{\nu^2}{1/\beta-2\nu}(\mathbf{r}_1-\mathbf{r}_2)^2}
  \phi(\mathbf{r}_{1})\phi(\mathbf{r}_{2})\\
& \propto 
  e^{-{(\mathbf{r}_1-\mathbf{r}_2)^2}/{2\sigma^{2}}}
  \phi(\mathbf{r}_{1})\phi(\mathbf{r}_{2}).\\
  \end{split}
\end{equation}
With parameter $\nu=0.22$ fm$^{-2}$ and optimized parameter $\beta=2.22$ fm$^2$,
the width $\sigma$ in the Gaussian term of Eq.~(\ref{eq:exact}) has very small
value of $\sigma=0.33$ fm, which confirms the contact of two nucleons in the
$NN$ pair as concluded from Eq.~(\ref{eq:simplified}).

Our finding of the contact representation of short range correlation reminds us
the recent study of the generalized contact theory by Weiss {\it et al.}
\cite{weiss18}.  They write the nuclear wave function in a factorized wave
function in the relative and center-of-mass coordinates of two nucleons in short
distance.  In our results, we find a simple factorized wave function to express
very high momentum components of short range correlation.
\begin{equation}
\psi(\mathbf{r_1},\mathbf{r_2})\propto
\delta(\mathbf{r_1}-\mathbf{r_2})
e^{-\frac{\nu}{2}(\mathbf{r_1}-\mathbf{r_2})^2}
e^{-\frac{\nu}{2}(\mathbf{r_1}+\mathbf{r_2})^2}
\end{equation}
It is interesting to note that the very high momentum components of the short range
correlation should be expressed by the simple contact form and the intermediate
range should be represented by the sum of medium range momentum components.

These findings significantly simplify the corresponding numerical calculations
in HMAMD framework.  In our future works, we will study the contact
correlation using the new wave function with the AV8$^\prime$ interaction,
where both the tensor interaction and short-range repulsion are included.

\section{Conclusion}
\label{sec:conclusion}
In conclusion, we found a contact representation of high momentum components of
short range correlation between nucleons through variation in the HMAMD
framework.  This contact wave function can be expressed in the HMAMD formalism
by a single Gaussian weighted AMD wave function with relative imaginary shift
components of Gaussian wave packets representing the momentum. We perform the
calculations of $^3$H with AV4$^\prime$ central potential using explicit
description of this new contact wave function and compare the results with those
of our previous works, in which the short-range correlation is not assumed in
any specific form in the basis states. We confirm the equivalent results
between two methods. It is found that our new formulation is very effective in
describing short range correlation, where a very rapid convergence is obtained in
the numerical calculation and the total basis number is reduced to almost the half
of the previous works. The present new understanding of the contact correlation
is expected to be very beneficial for future {\it ab initio} investigations of
nuclear systems.

\begin{acknowledgments}
The author would like to thank Mr. Wan for fruitful discussions. This work is
supported by the National Natural Science Foundation of China (grant nos
11535004, 11375086, 11120101005, 11175085, 11235001, 11761161001), by the
National Major State Basic Research and Development of China (grant nos
2016YFE0129300, 2018YFA0404403), by the Science and Technology Development Fund
of Macau under grant no. 008/2017/AFJ, and by the JSPS KAKENHI Grants No.
JP18K03660, and No. JP16K05351. M.L. acknowledges the support from the RCNP
theoretical group and the Yozo Nogami Research Encouragement Funding.
\end{acknowledgments}

\end{document}